\documentclass[twocolumn,preprintnumbers,amsmath,amssymb]{revtex4-1}

\usepackage{ulem}
\usepackage{graphicx}
\usepackage{amsmath}

\newcommand{\commentoutA}[1]{}

\begin{document}

\preprint{LA-UR-23-22053}

\title{Spin-Polarized Extended Lagrangian Born-Oppenheimer Molecular Dynamics}

\author{Yu Zhang, Marc J. Cawkwell, Christian F. A. Negre, Romain Perriot, Anders M. N. Niklasson}
\email{amn@lanl.gov}
\affiliation{Theoretical Division, Los Alamos National Laboratory, Los Alamos, New Mexico 87545}

\date{\today}

\begin{abstract}
We present a generalization of Extended Lagrangian Born-Oppenheimer molecular dynamics  [Phys.\ Rev.\ Lett.\ {\bf 100}, 123004 (2008); Eur. Phys. J. B {\bf 94}, 164 (2021)] that also includes the electronic spin-degrees of freedom as extended dynamical variables.  To integrate the combined spin and charge degrees of freedom, we use a preconditioned low-rank Krylov subspace approximation. Our approach is demonstrated for  quantum-mechanical molecular dynamics simulations of iron, using spin-polarized self-consistent charge density functional tight-binding theory. We also show how the low-rank Krylov subspace approximation can be used to accelerate the self-consistent field convergence.
\end{abstract}

\maketitle

\section{Introduction}

In quantum-mechanical Born-Oppenheimer molecular dynamics (QMD) simulations, forces acting on the atoms are determined from the fully relaxed electronic ground state in each MD time step, using Hartree-Fock  or density functional theory \cite{Roothaan,RMcWeeny60,hohen,RParr89,RMDreizler90}. However, iterative optimization is required to find a self-consistent field (SCF) solution for the electronic ground state due to the non-linear electron interactions. Insufficient convergence in the iterative optimization can result in inaccurate forces, which can invalidate the simulation. The iterative SCF optimization is usually the most computationally demanding task in QMD simulations, limiting the ability to study large systems under long simulation times. Furthermore, because of the non-linearities, QMD simulations can be sensitive to numerical approximations. Combining QMD simulations with linear scaling electronic structure theory using numerically thresholded sparse matrix algebra or with some approximate divide and conquer approach may therefore lead to convergence or stability problems \cite{MCawkwell12}. The same problems may also occur when calculations are performed in low-precision floating-point arithmetics using modern AI-hardware \cite{JFinkelstein21,JFinkelstein21B}.

To reduce the computational overhead of the self-consistent electronic ground-state optimization, several techniques have been developed \cite{TAArias92,PPulay04,JMHerbert05,ANiklasson06,TDKuhne07,JFang16}. One of these is the Car and Parrinello framework \cite{RCar85,DRemler90,GPastore91,FBornemann98,DMarx00,MTuckerman02,JHutter12}, which avoids charge relaxation by introducing the electronic degrees of freedom as classical dynamical field variables within an extended Lagrangian formulation. The extended electronic degrees closely follow the exact electronic ground  state by including orthonormality (or idempotency) constraints and a small fictitious electron mass parameter. As a result, the iterative ground state optimization is avoided. Car-Parrinello molecular dynamics (CPMD) simulations are effective as long as there is an electronic gap, and the electronic mass parameter is small enough to  guarantee an adiabatic separation between the slower nuclear motion and faster electronic evolution. However, in the limit of a small electronic gap, the electron mass parameter must be reduced to ensure that this separation is sufficient. This requires a short integration time step in order to capture the high frequency of the light and fast electronic motion, limiting the efficiency of CPMD simulations.

Recently, Bonella and co-workers proposed a mass-zero constrained molecular dynamics scheme that introduces exact adiabatic constraints on the electronic degrees of freedom in CPMD, enforced by Lagrange multipliers \cite{ACoretti18,SBonella20,ACoretti20}. The scheme provides a general theoretical framework for simulations of a broad range of problems with various types of electronic degrees of freedom, including wave functions, charge densities, or density matrices, similar to CPMD. The Lagrange multipliers need to be computed through an iterative procedure that needs to be well converged. However, the integration time step can be longer -- of the same order as in regular classical MD. This elegant approach provides an efficient alternative to the original formulation of CPMD.

Another method to avoid the non-linear SCF problem in QMD simulations is extended Lagrangian Born-Oppenheimer molecular dynamics (XL-BOMD) \cite{ANiklasson08,PSteneteg10,GZheng11,MCawkwell12,JHutter12,LLin14,MArita14,PSouvatzis14,ANiklasson14,KNomura15,AAlbaugh15,ANiklasson17,JBjorgaard18,ANiklasson21,ANiklasson21b}. XL-BOMD is a technique that includes the electronic degrees of freedom as extended dynamical variables, similar to CPMD, but with a different extended Lagrangian. The equations of motion are derived in a mass-zero limit using an approximate shadow Born-Oppenheimer potential, and the time step can be chosen on the same order as regular Born-Oppenheimer molecular dynamics (BOMD). However, unlike BOMD, XL-BOMD does not require an iterative SCF optimization procedure.
Instead, a kernel has to be calculated that determines the generalized metric tensor of the extended harmonic oscillator that drives the fictitious electron dynamics. A direct exact construction
of this kernel in each time step would be quite expensive. However, for most non-reactive systems with an electronic gap, a simple constant kernel approximations can be used that appears similar to a preconditioner. For more demanding problems, e.g.\ systems with charge sloshing 
and a small electronic energy gap (or a closing gap), this is not always sufficient. 
Recent techniques have been proposed to approximate the kernel using preconditioned Krylov subspace approximations, enabling XL-BOMD simulations of chemically unstable systems undergoing exothermic reactions \cite{ANiklasson20,ANiklasson20b,ANiklasson21b,ANiklasson23,CNegre23}. 

In this article, we aim to extend XL-BOMD and the preconditioned Krylov subspace approximation techniques to spin-polarized systems  with collinear spin degrees of freedom. So far, XL-BOMD has only been applied to non-magnetic systems. Recently, Das and Gavini proposed a spin-polarized version of the low-rank Krylov subspace approximation to accelerate the convergence of the iterative SCF optimization procedure, which demonstrated excellent performance compared to other methods \cite{VGavini22}.
Our goal in this article is to generalize both XL-BOMD and the preconditioned Krylov subspace approximation techniques to include also spin-polaraized systems with collinear spin degrees of freedom. By extending XL-BOMD to spin-polarized systems we will show how efficient SCF-free QMD simulations can be performed even for ferromagnetic metals such as Fe. Spin-polarized QMD simulations of metals can be quite challenging using regular Born-Oppenheimer methods. With spin-polarized XL-BOMD (sXL-BOMD) many of those challenges can be avoided.

We first define the spin-polarized extended Lagrangian for collinear spin-polarized density functional theory (DFT). We discuss the approximate shadow Born-Oppenheimer potential used in the Lagrangian in terms of both general Hohenberg-Kohn and Kohn-Sham DFT, as well as the equations of motion and their integration. We present the spin-polarized kernel and how it can be represented through a low-rank preconditioned Krylov subspace approximation.

To demonstrate the theory, we then perform sXL-BOMD simulations of Fe using spin-polarized self-consistent charge density functional tight-binding theory (SCC-DFTB) \cite{MElstner98,MFinnis98,TFrauenheim00,MGaus11,BAradi15,BHourahine20}. The simulation results show the effectiveness of spin-polarized XL-BOMD for simulating spin-polarized systems, including ferromagnetic metals, without the need for an iterative SCF optimization procedure.

\section{Spin-Polarized Extended Lagrangian Born-Oppenheimer Molecular dynamics}\label{sXLBOMD}

In this section we will introduce the generalization of XL-BOMD that also includes collinear spin-degrees of freedom as extended dynamical variables together with the atomic positions and velocities. First the theory is described using general Hohenberg-Kohn theory in sec.\ \ref{HK} and sec.\ \ref{ShadowHK} before introduce the Kohn-Sham formulation in sec.\ \ref{ShadowKS}. The dynamics is introduced through an extended Lagrangian in Eq.\ (\ref{SXL-BOMD}) using a shadow Born-Oppenheimer potential, defined in Eq.\ (\ref{Shadow}) or Eq.\ (\ref{ShadowU}), which in an adiabatic limit leads to the equations of motion in Eq.\ (\ref{EqMot}), where the kernel can be approximated using the preconditioned low-rank approximation in Eq.\ (\ref{mIJacobian}). These equations summarize the main part of the theory in this section.

\subsection{The Extended Born-Oppenheimer Lagrangian}\label{HK}

A spin-polarized extended Lagrangian Born-Oppenheimer molecular dynamics (sXL-BOMD) can be defined through an extended Lagrangian,
\begin{equation}\label{SXL-BOMD}\begin{array}{l}
{\displaystyle {\cal L}({\bf R},{\bf {\dot R}}, {\bf n}, {\bf \dot n}) = }\\
~~\\
{\displaystyle  = \frac{1}{2} \sum_I M_I {\dot R}_I^2 -{\cal U}_{\rm BO}({\bf R},{\bf n}) + \frac{\mu}{2} \sum_{\sigma } \int {\dot n_{\sigma }}^2({\bf r}) d{\bf r}}\\
~~\\
{\displaystyle  - \frac{\mu \omega^2}{2} \sum_{\sigma ,\sigma' } \iint f_\sigma [{\bf n}]({\bf r}) 
T_{\sigma , \sigma'  } ({\bf r},{\bf r'})  f_{\sigma '} [{\bf n}]({\bf r'}) d{\bf r} d{\bf r'} }.
\end{array}
\end{equation}
where $f_\sigma [{\bf n}]({\bf r})$ is a spin-dependent residual function,
\begin{equation}
f_\sigma [{\bf n}]({\bf r}) = \left(q_{\sigma}[{\bf n} ]({\bf r}) - n_{\sigma }({\bf r}) \right).
\end{equation}
Here ${\cal U}_{\rm BO}({\bf R},{\bf n})$ is a shadow Born-Oppenheimer potential that approximates the exact regular Born-Oppenheimer potential, which is defined below in Eqs.\ (\ref{Shadow})-(\ref{LinF}).
The collinear spin density is given in terms of a spin up and down density,
\begin{equation} \label{TwoComp}
{\displaystyle {\bf n} = \left[ \begin{array}{l} n_{\uparrow } ({\bf r})\\ n_{\downarrow}({\bf r})\end{array}\right], ~~
{\bf q}[{\bf n}] = \left[ \begin{array}{l} q_{\uparrow} [{\bf n}]({\bf r})\\ q_{\downarrow} [{\bf n}]({\bf r})\end{array}\right]},
\end{equation}
where $\sigma = \{\uparrow, \downarrow\} $ denotes the two spin channels. 
The dynamical variables of the Lagrangian are the nuclear positions, ${\bf R} = \{{\bf R}_I\}$ (${\bf R}_I\ = \{R_{Ix},R_{Iy},R_{Iz}\}$), their
velocities, ${\bf {\dot R}} = \{{\bf \dot R}_I\}$,  the spin density ${\bf n}({\bf r})$, and its time derivative ${\bf \dot n}({\bf r})$.
The extended Harmonic oscillator part of the extended Lagrangian includes $\mu$, 
which is a fictitious mass parameter of the electronic degrees of freedom and $\omega$, which
is the frequency of the extended harmonic oscillator, where $n_\sigma ({\bf r})$ oscillates around the variationally optimized ground state density, 
$q_\sigma [{\bf n} ]({\bf r})$ (given in Eqs.\ (\ref{q_min}) and (\ref{q_min_x}) below) that defines the shadow Born-Oppenheimer potential, ${\cal U}_{\rm BO}({\bf R},{\bf n})$.
The extended harmonic oscillator includes a symmetric and positive definite metric tensor,
\begin{equation}
{\displaystyle T_{\sigma \sigma'}({\bf r,r'}) = \sum_{\sigma''} \int K^\dagger_{\sigma \sigma''} ({\bf r},{\bf r''})K_{\sigma'' \sigma'}({\bf r''},{\bf r'}) d{\bf r''}}, 
\end{equation}
which is determined by a kernel, $K_{\sigma \sigma'}({\bf r},{\bf r'})$. We will later define this kernel as the inverse
of the Jacobian of the spin-density residual functional, ${\bf f(n)}  = {\bf q[{\bf n} ]({\bf r}) - n({\bf r})}$, which provides
a particularly efficient choice. 

Apart from the splitting of the electronic degrees of freedom into two spin channels, the formalism above follows previous formulations of XL-BOMD  \cite{ANiklasson14,ANiklasson17,ANiklasson20,ANiklasson21b}.

\subsection{Shadow Born-Oppenheimer Potential}\label{ShadowHK}

In regular BOMD based on Hohenberg-Kohn 
spin density functional theory \cite{hohen,RParr89,RMDreizler90} the potential energy surface is given by
\begin{equation}\label{U_min}
U_{\rm BO}({\bf R}) = F[{\boldsymbol \rho^{\rm min}}] + \int v({\bf R, r}) \rho^{\rm min}({\bf r}) 
d{\bf r} + V_{nn}({\bf R}).
\end{equation}
The last term of $U_{\rm BO}({\bf R})$ in Eq.\ (\ref{U_min}) is the nuclear ion repulsion and
${\boldsymbol \rho}^{\rm min}({\bf r}) = [\rho^{\rm min}_\uparrow, \rho^{\rm min}_\downarrow]^T$
is the relaxed electronic ground state spin density, where
$\rho^{\rm min}({\bf r}) = \rho^{\rm min}_\uparrow ({\bf r}) + \rho^{\rm min}_\downarrow ({\bf r})$. 
This ground state spin density is given from a constrained minimization over all normalized and physically relevant (or $v$-representable) spin densities, 
${\boldsymbol \rho}({\bf r})$,
of the universal spin-density functional, $F[{\boldsymbol \rho}]$, including an external potential, 
$v({\bf R},{\bf r})$, which here is assumed to be spin-independent and purely electrostatic, i.e.\
\begin{equation} \label{rho_min} \begin{array}{l}
{\displaystyle {\boldsymbol \rho}^{\rm min}({\bf r}) = 
\arg \min_{{\boldsymbol \rho} } \left\{ F[{\boldsymbol \rho}]
+ \int v({\bf R, r}) \rho({\bf r}) d{\bf r} \right. }\\
~~\\
{\displaystyle ~~~~~~~~~~~~~~ \left. \left \vert ~\int \rho({\bf r}) d{\bf r} = N_e \right. \right\}}.
\end{array}
\end{equation}

In our spin-polarized generalization of XL-BOMD, as defined by the extended Lagrangian in Eq.\ (\ref{SXL-BOMD}), the potential is given by an approximate \textit{shadow} potential energy surface, 
\begin{equation}\label{Shadow} 
{\displaystyle {\cal U}_{\rm BO}({\bf R},{\bf n}) = {\cal F}[{\bf q}[{\bf n}],{\bf n}]
+ \int v({\bf R},{\bf r}) {\bf q}[{\bf n}]({\bf r}) d{\bf r} + V_{nn}({\bf R})}.
\end{equation}
Here  ${\bf q}[{\bf n}]({\bf r}) = [q_\uparrow[{\bf n}], q_\downarrow[{\bf n}]]^T$ is 
the relaxed stationary electronic spin density ground state
of the shadow potential, which is determined by the constrained minimization, 
\begin{equation}\label{q_min} \begin{array}{l}
{\displaystyle {\bf q}[{\bf n}]({\bf r}) = \arg \min_{{\boldsymbol \rho} } \left\{ {\cal F}[{\boldsymbol \rho},{\bf n}] + \int v({\bf R, r}) \rho({\bf r}) d{\bf r} \right. }\\
~~\\
{\displaystyle ~~~~~~~~~~~~~~ \left. \left \vert ~ \int \rho({\bf r}) d{\bf r} = N_e \right. \right\}},
\end{array}
\end{equation}
or, more generally, as the variationally stationary solution that satisfies:
\begin{equation}\label{q_min_x}\begin{array}{l}
{\displaystyle \frac{\delta}{\delta \rho_\sigma} \left\{ {\cal F}[{\boldsymbol \rho}, {\bf n}] + \int v({\bf R, r}) \rho({\bf r}) d{\bf r}\right\}_{{\boldsymbol \rho} \in v}= 0, ~~ \sigma = \{\uparrow,\downarrow\} }\\
~~\\
{\displaystyle \int \rho({\bf r}) d{\bf r} = \int \left( \rho_{\uparrow} ({\bf r}) + \rho_{\downarrow}({\bf r}) \right)  d{\bf r} = N_e }.
\end{array}
\end{equation}
Here ${\cal F}[{\boldsymbol \rho}, {\bf n}]$ is an approximate, linearized universal shadow energy functional,
\begin{equation}\label{LinF} \begin{array}{l}
{\displaystyle {\cal F}[{\boldsymbol \rho},{\bf n} ] = F[{\bf n} ] 
+ \sum_\sigma \int \left. \frac{\delta F[{\boldsymbol \rho}]}{\delta \rho_\sigma ({\bf r})} \right \vert_{{\bf \rho}_\sigma = n_\sigma}
\left( \rho_\sigma ({\bf r}) - n_\sigma({\bf r})\right) d{\bf r}}\\
~~\\
{\displaystyle ~~~~~~~~~ = F[{\boldsymbol \rho}] + {\cal O}[\vert {\boldsymbol \rho} - {\bf n} \vert^2]},
\end{array}
\end{equation}
which is linearized around the electron spin-density field variable ${\bf n}({\bf r})$. If ${\bf n}({\bf r})$ is close to the exact
Born-Oppenheimer ground state spin density, ${\boldsymbol \rho}^{\rm min}({\bf r})$, the error in the constrained optimized solution, 
${\bf q}{\bf [n]}$, of the linearized shadow functional, which is given from the solution of Eq.\ (\ref{q_min_x}), is also small.
The major advantage over regular, direct Born-Oppenheimer molecular dynamics is that this optimization
can be performed in a single step, without any iterative SCF optimization. 
In Kohn-Sham DFT only a single construction and diagonalization of the effective single-particle Kohn-Sham Hamiltonian is necessary.
However, the electronic stationary state, ${\bf q[n]}({\bf r})$, 
of the shadow potential, ${\cal U}_{\rm BO}({\bf R},{\bf n})$, 
is only an ${\bf n}$-dependent approximation of the true Born-Oppenheimer density, ${\boldsymbol \rho}^{\rm min}({\bf r})$. The error will depend on the square of the
residual, ${\bf f[n]} = ({\bf q}{\bf [n] - n)}$, which oscillates as ${\bf n}({\bf r})$ 
evolves through the harmonic oscillator centered around ${\bf q}{\bf [n]}({\bf r})$
in Eq.\ (\ref{SXL-BOMD}). 

The shadow Born-Oppenheimer potential, ${\cal U}_{\rm BO}({\bf R,n})$,
can be calculated without requiring any iterative SCF optimization. 
However, ${\cal U}_{\rm BO}({\bf R,n})$ presented here is meaningful only
in the context of sXL-BOMD, where the spin density, ${\bf n}$, is included as a time-dependent dynamical field variable, ${\bf n}(t)$.
Only then is it possible to use ${\cal U}_{\rm BO}({\bf R,n})$ for the calculation of the interatomic forces. 
This is in contrast to a Harris-Foulkes functional \cite{JHarris85,MFoulkes89}, 
which equals ${\cal U}_{\rm BO}({\bf R,n})$ for Kohn-Sham
DFT in the static, non-dynamical case. In a Harris-Foulkes functional ${\bf n}$ would be some approximate ground state input density, for example, a
superposition of atom centered spin densities \cite{ANiklasson14,ANiklasson17,ANiklasson23}. 
The Harris-Foulkes energy functional can then be used to approximate
the ground state energy, but it is not accurate for the evaluation of forces.
Thus, ${\cal U}_{\rm BO}({\bf R,n})$ in sXL-BOMD can be seen as a generalized extension of the Harris-Foulkes energy functional that is applicable also to force calculations. Conceptually the shadow Born-Oppenheimer potential is therefore different from a Harris-Foulkes functional. ${\cal U}_{\rm BO}({\bf R,n})$ is better understood from the perspective of a backward error analysis or a shadow Hamiltonian approach 
\cite{HYoshida90,SToxvaerd94,GJason00,SToxvaerd12,ShadowHamiltonian}. 
Instead of estimating approximate energies
for the exact Born-Oppenheimer potential, we calculate both the exact energies and the exact forces (at little cost), 
but for our approximate shadow Born-Oppenheimer potential energy surface. sXL-BOMD can therefore be seen as a shadow molecular dynamics scheme for a non-linear SCF model. The concept of shadow molecular dynamics has been very successful in the design of symplectic or geometric integration schemes. These integration schemes often combine a high accuracy with outstanding long-term stability. Physically important properties, such as a conserved phase-space area and the total energy conservation, can be controlled rigorously. Our proposed sXL-BOMD has many similar qualities.

\subsection{Shadow Born-Oppenheimer potential for spin-polarized Kohn-Sham DFT with fractional occupation numbers}\label{ShadowKS}

In Kohn-Sham DFT it is assumed that
the electron density, ${\boldsymbol \rho}({\bf r})$, is given from single-particle orbitals, $\{\Phi_k^\sigma\}$,
\begin{equation}
{\boldsymbol \rho}({\bf r}) = \sum_{k,\sigma} f^\sigma_k \vert \Phi^\sigma_k({\bf r}) \vert^2,
\end{equation}
where ${\bf f} = \{f^\sigma_k\}$ are the occupation numbers, $f^\sigma_k \in [0,1]$.
The Kohn-Sham (KS) spin-density (free-energy) functional is then given by the ansatz,
\begin{align} \label{KS_Func} 
F_{\rm KS}[{\boldsymbol \rho}] =& -\frac{1}{2} \sum_{k,\sigma} f^\sigma_k \langle \Phi^\sigma_k({\bf r}) \vert \nabla^2 \vert \Phi^\sigma_k({\bf r}) \rangle \nonumber \\
&+ \frac{1}{2}\iint \frac{{\rho}({\bf r}) { \rho}({\bf r'}) }{\vert {\bf r-r'}\vert } + E_{\rm xc}[{\boldsymbol \rho}]
- T_e {\cal S}[{\bf f}]. 
\end{align}
The first term is the kinetic energy term, the second term is the Hartree term, and the third term is the exchange-correlation
energy, $E_{\rm xc}[{\boldsymbol \rho}]$,  which more typically but equivalently, is given in terms of the net charge density, 
$\rho({\bf r}) = \rho_\uparrow({\bf r}) + \rho_\downarrow({\bf r})$, and the spin density, 
$m({\bf r}) = \rho_\uparrow({\bf r}) - \rho_\downarrow({\bf r})$. 
The last term is the product between the electronic temperatures, $T_e$, and the electronic mean-field entropy term,
\begin{equation}
{\cal S}[{\bf f}] = -k_B \sum_{k,\sigma} f^\sigma_k \ln (f^\sigma_k) + (1-f^\sigma_k)\ln (1-f^\sigma_k),
\end{equation}
where $k_B$ is the Boltzmann constant. 

The linearized approximate shadow energy functional, as in Eq.\ (\ref{LinF}), is then given by
\begin{align}\label{FreeShadowFunc}
&{\cal F}_{\rm KS}[{\boldsymbol \rho},{\bf n}] = -\frac{1}{2} \sum_{k,\sigma} f^\sigma_k \langle \Phi^\sigma_k({\bf r}) \vert 
\nabla^2 \vert \Phi^\sigma_k({\bf r}) \rangle \nonumber \\
& ~~~~~+ \frac{1}{2}\iint \frac{\left(2{\rho}({\bf r}) - n({\bf r})\right){ n}({\bf r'}) }{\vert {\bf r-r'}\vert } + E_{\rm xc}[{\bf n}] \nonumber \\
& ~~~~~+ \sum_\sigma \int v^\sigma_{\rm xc}[{\bf n}]({\bf r})\left(\rho_\sigma({\bf r}) - n_\sigma({\bf r})\right) d{\bf r} \nonumber\\
& ~~~~~- T_e {\cal S}[{\bf f}], 
\end{align}
with the exchange-correlation potential, 
\begin{equation}
v^\sigma_{\rm xc}[{\bf n}]({\bf r}) = \left. \frac{\delta E_{\rm xc}[{\boldsymbol \rho}({\bf r})]}{\delta \rho_\sigma({\bf r})}\right \vert_{{\boldsymbol \rho} = {\bf n}}.
\end{equation}
The single-particle orbitals can be approximated fairly generally through an ansatz using a linear combination of a finite set of atomic-orbital-like basis functions,
\begin{equation}
\Phi_k^\sigma({\bf r}) = \sum_i c^{(k)}_{\sigma,i} \phi_i({\bf r}),
\end{equation}
such that
\begin{align}\label{rho_from_D}
 \rho_\sigma({\bf r}) &= \sum_{k, \sigma} f^\sigma_k \vert \Phi_k^\sigma({\bf r})\vert^2 
= \sum_{i,j} \left(\sum_k f^\sigma_k c^{(k)}_{\sigma,i} c^{(k)}_{\sigma,j}\right) \phi_i({\bf r})\phi_j({\bf r})\nonumber \\
& = \sum_{ i,j} D^\sigma_{ij} \phi_i({\bf r})\phi_j({\bf r}),
\end{align}
which defines the spin-dependent density matrix, $D^\sigma_{ij} = \sum_k f^\sigma_k c^{(k)}_{\sigma,i} c^{(k)}_{\sigma,j}$. For simplicity, the atomic orbitals, $\{ \phi_i({\bf r}) \}$, are assumed to be real atom-centered functions, but
the formalism can easily be generalized to complex basis sets. 
Instead of working with the basis function expansion coefficients, $\{c^{(k)}_{\sigma,i}\}$, we may use
the elements of the density matrix, $D^\sigma$, as the free parameters that determine the ground state spin density.

By performing a constrained minimization over the spin-dependent density matrix elements, $\{D^\sigma_{ij}\}$, we can then find the lowest stationary solution as in Eq.\ (\ref{q_min_x}) that defines our shadow Born-Oppenheimer potential i.e.,
\begin{align}\label{ShadowU}
 {\cal U}({\bf R,n}) = &\min_{D^\sigma} \left\{ {\cal F}_{\rm KS}[{\boldsymbol \rho},{\bf n}] + \int v({\bf R,r}) \rho({\bf r}) d{\bf r} \right. \nonumber \\
& \left. -\mu \left( \sum_\sigma \int \rho_\sigma({\bf r}) d{\bf r} - N_e \right) \right\}.
\end{align}
The stationary ground-state density matrix is given by
\begin{equation}\label{D_min} \begin{array}{l}
{\displaystyle {D^{\sigma}}[{\bf n}]^\perp = \left[e^{\left({H^\sigma}[{\bf n}]^\perp - \mu I)\right)}+I\right]^{-1}   }
\end{array}
\end{equation}
where the chemical potential, $\mu$, is chosen such that the total number of electron $\sum_\sigma \int \rho_\sigma ({\bf r}) d{\bf r} = N_e$ is correct.
Here $I$ is the identity matrix, ${H^\sigma}[{\bf n}]^\perp = Z^T{H^\sigma}[{\bf n}]Z$ and $D^{\sigma}[{\bf n}] = Z{D^{\sigma}}[{\bf n}]^\perp Z^T $, and
the Kohn-Sham Hamiltonian matrix, $H^\sigma[{\bf n}]$, has matrix elements
\begin{align}
 H^\sigma_{ij}[{\bf n}] & =\int \phi_i({\bf r}) \left(-\frac{\nabla^2}{2} + V_{\rm H}[n]({\bf r}) \right) \phi_j({\bf r}) d{\bf r} \nonumber \\
&+ \int \phi_i({\bf r}) \left( + v^\sigma_{\rm xc}[{\bf n}]({\bf r})   + v({\bf R,r}) \right) \phi_j({\bf r}) d{\bf r}.
\end{align}
The overlap matrix, $S$, has matrix elements
\begin{equation}
S_{ij} = \int \phi_i({\bf r}) \phi_j({\bf r}) d{\bf r},
\end{equation}
and the congruence transformation matrix, $Z$, is the inverse square root of $S$, or more generally
$Z$ is defined such that 
\begin{equation}
Z^TSZ = I.
\end{equation}
 The $n$-dependent ground state spin-charge, ${\bf q}[{\bf n}]({\bf r})$, in Eq.\ (\ref{q_min}),
is then given by
\begin{equation}
{\displaystyle { q}_\sigma [{\bf n}]({\bf r}) 
= \sum_{i,j} D^\sigma_{ij}[{\bf n}] \phi_i({\bf r})\phi_j({\bf r})},
\end{equation}
with the spin-dependent density matrix, $D^\sigma$, given by Eq.\ (\ref{D_min}). This density matrix, $D^\sigma$, can be constructed in a single direct step without any iterations -- no self-consistency is required.
The ground state spin density, ${\bf q}[{\bf n}]({\bf r})$, then defines the spin-dependent shadow Kohn-Sham Born-Oppenheimer potential, ${\cal U}_{BO}(\mathbf{R},\mathbf{n})$, as in Eq.\ (\ref{Shadow}), which is used in the extended Lagrangian in Eq.\ (\ref{SXL-BOMD}) that defines our sXL-BOMD.

\subsection{Equations of Motion}

The equations of motion for spin-polarized XL-BOMD defined by the Lagrangian in Eq.\ (\ref{SXL-BOMD}) 
can be derived from Euler-Lagrange equations in an adiabatic limit, where
$\omega \rightarrow \infty$ and $\mu \rightarrow 0$, such that $\omega \mu = {\rm constant}$ \cite{ANiklasson14,ANiklasson17,ANiklasson21b}.
This limit is a classical adiabatic approximation for the extended electronic degrees of freedom.
The adiabatic limit is similar to the Born-Oppenheimer approximation \cite{DMarx00}. We simply use the same underlying assumption as for the Born-Oppenheimer approximation, where the electronic degrees of freedom are fast compared to the slower nuclear motion. This approximation is applied to the classical extended electronic degrees of freedom, which are propagated by the extended Harmonic oscillator with a time scale determined by the oscillator frequency $\omega$.
It can be shown that this adiabatic approximation is justified under normal conditions, i.e.\ when we use integration time steps as in regular BOMD simulations, and that it is system independent \cite{ANiklasson17,ANiklasson21b}. In continuous time the adiabatic limit is exact. For Kohn-Sham density functional theory the equations of motion in this classical Born-Oppenheimer adiabatic limit are
\begin{equation}\label{EqMot}\begin{array}{l}
{\displaystyle M_I {\ddot R}_{I \alpha} = \left.-\frac{\partial {\cal U}_{\rm BO}({\bf R,n})}{\partial R_{I\alpha}}\right\vert_{\bf n}}, ~~ \alpha = x,y,z\\
~~\\
{\displaystyle {\ddot n}_\sigma ({\bf r}) = -\omega^2 \sum_{\sigma'} 
\int K_{\rm \sigma \sigma'}({\bf r},{\bf r'})f_{\sigma'} [{\bf n}]({\bf r'})d{\bf r'}},
\end{array}
\end{equation}
if the kernel $K_{\sigma \sigma'}({\bf r},{\bf r'})$ is defined as the inverse Jacobian of the residual function,
$ f_\sigma [{\bf n}]({\bf r}) = {q_\sigma [n]}({\bf r})-{n_\sigma}({\bf r})$, i.e.\ such that
\begin{align}\label{Kernel_1}
& \sum_{\sigma'}\int K_{\sigma \sigma'}({\bf r},{\bf r'}) \left(\frac{\delta q_{\sigma'}[{\bf n}]({\bf r'})}{\delta n_{\sigma''}({\bf r''})}
- \frac{\delta n_{\sigma'}({\bf r'})}{\delta n_{\sigma''}({\bf r''})}\right) d{\bf r'} \nonumber \\
&~~~~~~~~~= \delta_{\sigma,\sigma''} \times \delta({\bf r-r''}).
\end{align}

The first set of equations for the nuclear coordinates in Eq.\ (\ref{EqMot}), with the partial
derivative of the shadow potential calculated under constant spin density ${\bf n}$, looks similar
to regular Born-Oppenheimer molecular dynamics. No $(\delta {\cal U}_{\rm BO}/\delta {\rho_\sigma})(\partial {\bf \rho_\sigma}/\partial R_I)$
terms are needed, because $\delta {\cal U}_{\rm BO}/\delta {\rho_\sigma} = 0$ at $ {\boldsymbol \rho} = {\bf q}[{\bf n}]({\bf r})$, which 
is the constrained stationary ground sate solution to the functional that defines ${\cal U}_{\rm BO}({\bf R},{\bf n})$ in Eq.\ (\ref{q_min}).  
The forces are based on the shadow
potential energy surface, ${\cal U}_{\rm BO}({\bf R,n})$, which can be calculated directly in a single step without any
potential convergence problems.
This is in contrast to regular direct Born-Oppenheimer molecular dynamics, where the forces are
calculated for the optimized electronic ground state relying on the Hellman-Feynman theorem. 
This requires a costly iterative SCF optimization which in practice never is complete.

The inverse Jacobian kernel, $K_{\sigma,\sigma'}({\bf r},{\bf r'})$, in the second equation of motion in Eq.\ (\ref{EqMot}), which is defined in Eq.\ (\ref{Kernel_1}), appears as a Newton optimization step acting
on the residual such that $n_{\sigma} ({\bf r})$ behaves as if it would oscillate around a much closer approximation
to the exact Born-Oppenheimer ground state density, $\rho^{\rm min}_{\sigma}({\bf r})$, \cite{ANiklasson21b,ANiklasson23}.
The definition of the kernel as the inverse Jacobian of the residual function in Eq.\ (\ref{Kernel_1}) therefore plays
a dual role: it makes the electronic equations of motion simple in the adiabatic limit and it serves as
a stabilizer that improves the accuracy by making the spin density ${\bf n}({\bf r})$ evolve more closely to the exact
Born-Oppenheimer ground state density, ${\boldsymbol \rho}^{\rm min}({\bf r})$. This also means that ${\bf q[n]}({\bf r})$ will
be close to ${\boldsymbol \rho}^{\rm min}({\bf r})$ and that the residual spin density function, ${\bf f}[{\bf n}]({\bf r}) = {\bf q[n]}({\bf r})-{\bf n}({\bf r})$, 
stays small during an XL-BOMD simulation.

\subsection{Constant of motion}

The constant of motion in the adiabatic limit is given by the total energy expression
\begin{equation}\label{ShadowH1}
{\displaystyle {\cal E}^{\rm tot}_{\rm XBO} = \frac{1}{2} \sum_I M_I {\dot R}_I^2 + {\cal U}_{\rm BO}({\bf R},{\bf n})},
\end{equation}
which closely follows the exact Born-Oppenheimer constant of motion,
\begin{equation}\label{BOH}
E^{\rm tot}_{\rm BO} = \frac{1}{2}\sum_I M_I {\dot R}_I^2 + U_{\rm BO}({\bf R}).
\end{equation}

The error in the sampling of the potential energy surface is of second order in the residual, i.e.\
$\| U_{\rm BO} ({\bf R})- {\cal U}_{\rm BO}({\bf R,n})\| \propto \|{\bf q[n]-n}\|^2$,
which can be shown to scale to fourth-order in the size of the integration time step, $\delta t$, i.e.\ as ${\cal O}(\delta t^4)$,
if we use a Verlet-based integration scheme \cite{ANiklasson17,ANiklasson21b}. This means that $\| U_{\rm BO} ({\bf R})- {\cal U}_{\rm BO}({\bf R,n})\| \propto \delta t^4$.


\subsection{Integrating the equations of motion}

There are several alternative techniques to characterize and integrate the equations of motion, in Eq.\ (\ref{EqMot}), 
\cite{AOdell09,AOdell11,AAlbaugh15,VVitale17,AAlbaugh15,AAlbaugh17,AAlbaugh18,ILeven19}. The method that seems to be most efficient
for Kohn-Sham DFT or semi-empirical methods is
a modified leapfrog velocity Verlet integration scheme \cite{ANiklasson09,PSteneteg10,GZheng11,ANiklasson21b}.
If we apply this method for the integration
of the nuclear and electronic spin-density degrees of freedom, we get
\begin{align}\label{Integration}
& {{\bf \dot {\bf R}}_I}(t + \frac{\delta t}{2})  = {{\bf \dot {R}}_I}(t) + \frac{\delta t}{2} {{\bf \ddot {\bf R}}_I}(t)\\
&{{\bf R}_I}(t + \delta t) = {{\bf R}}_I(t) + \delta t {{\bf \dot {\bf R}}_I}(t + \frac{\delta t}{2})\\
& n_\sigma(t + \delta t) = 2n_\sigma(t) - n_\sigma(t - \delta t) + \delta t^2 {\ddot n}_\sigma(t) \nonumber \\
& ~~~~~~~~~~~~~~~~~+ \alpha \sum_{k = 0}^{k_{\rm max}} c_k n_\sigma(t-k \delta t)\\
&{{\bf \dot {\bf R}}_I}(t + \delta t) = {{\bf \dot {\bf R}}_I}(t + \frac{\delta t}{2})
+ \frac{\delta t}{2} {{\bf \ddot {\bf R}}_I}(t+\delta t).
\end{align}
The coefficients, $\alpha$ and $\{ c_k \}_{k = 0}^{k_{\rm max}}$,
as well as a dimensionless constant, $\kappa = \delta t^2 \omega^2$,
for various values of $k_{\rm max}$ are given in Ref.\ \cite{ANiklasson09}.
At the initial time step, $n_\sigma(t_0)$ and $n_\sigma(t_0-k\delta t)$ are all set to the
fully converged regular Born-Oppenheimer ground state density, $\rho^{\rm min}_\sigma({\bf r})$, at $t_0$.
A full regular SCF optimization is then required, but only at the first initial time step.
The $\alpha$-dependent term introduces a weak dissipation that keeps $n_\sigma(t)$ aligned
with the evolution of $q_\sigma[{\bf n}]$ and $\rho^{\rm min}_\sigma$.

In the integration we need to calculate ${{\ddot n}}_\sigma(t)$ in Eq.\ (\ref{EqMot}) in each time step. To do this we need 
to find a suitable approximation of the spin-dependent inverse Jacobian kernel $K_{\sigma,\sigma'}$. 
For example, we may use a fixed approximation of the kernel
based on a scaled delta function, or an exact full calculation of the kernel at the initial time step that is then used as a fixed approximation \cite{ANiklasson17}. 
For more challenging problems, including degenerate states in reactive chemical systems and metals, 
we can use a preconditioned low-rank Krylov subspace approximation \cite{ANiklasson20,DAKnoll04}
discussed below.

A key challenge in the integration of the equations of motion in Eq.\ (\ref{EqMot}) is to achieve a synchronization between
the evolution of the extended electronic degrees of freedom and the nuclear motion that determines the
exact regular Born-Oppenheimer electronic ground state. Without this synchronization, the extended electronic
degrees of freedom may drift away from the exact ground state solution and we then lose accuracy in linearization of the universal energy functional
around the extended electronic degrees of freedom.

\subsection{Approximating the inverse Jacobian kernel using a preconditioned Krylov subspace approximation}\label{Kernel}

The equation of motion for the electronic degrees of freedom in Eq.\ (\ref{EqMot})
can be simplified for a $2N$-finite dimensional space,
using a matrix notation, ${\bf q}[{\bf n}]:\mathbb{R}^{2N} \rightarrow \mathbb{R}^{2N}$ and ${\bf n} \in \mathbb{R}^{2N}$,
where the first $N$ components corresponds to the spin-up channel and the last $N$ components represent the spin-down channel.
In this case we may rewrite the electronic equation of motion in Eq.\ (\ref{EqMot}) in an algebraic matrix-vector form,
\begin{equation}\label{Eq_Mot2}
{\displaystyle {\bf \ddot n} = -\omega^2  {\bf K}\left({\bf q[n]-n}\right)},
\end{equation}
where the kernel ${\bf K} = \{K^{\sigma \sigma'}_{ij}\}$ is a $2N \times 2N$ matrix consisting of four $N\times N$ blocks
for the four different combinations of $\sigma$ and $\sigma'$.
The kernel ${\bf K}$ is the matrix inverse of the Jacobian matrix,
${\bf J} = \{J^{\sigma \sigma'}_{ij}\}$, i.e. ${\bf K} = {\bf J}^{-1}$, with matrix elements,
\begin{equation}\label{RegJacobian}
{\displaystyle J^{\sigma \sigma'}_{ij} = \frac{\partial f^{\sigma}_i({\bf n})}{\partial n^{\sigma'}_j}} 
\end{equation}
that are calculated from the partial derivatives of the spin-dependent residual function,
\begin{equation}\label{ResFunc}
{\displaystyle {\bf f}({\bf n}) = {\bf q[n]-n}, ~{\rm where}~ {\bf f}[{\bf n}]:\mathbb{R}^{2N}\rightarrow \mathbb{R}^{2N}}.
\end{equation}
We will use this simplified matrix-vector notation in the presentation of the low-rank preconditioned 
Krylov subspace approximation of the Kernel. The preconditioned formulation is based on an equivalent reformulation of Eq.\ (\ref{Eq_Mot2}), where
\begin{equation}\label{Eq_Mot3}
{\displaystyle {\bf \ddot n} = -\omega^2  \left({\bf K}_0{\bf J}\right)^{-1} {\bf K}_0\left({\bf q[n]-n}\right)}.
\end{equation}
In this reformulation we inserted an approximate kernel, ${\bf K}_0 \approx {\bf J}^{-1}$, as our preconditioner.

The standard definition of the Jacobian in Eq.\ (\ref{RegJacobian}) is based on partial derivatives with respect to the different
components of ${\bf n}$. This definition can be generalized for the preconditioned Jacobian, ${\bf K}_0{\bf J}$, using a more flexible form. We start by introducing a complete set of arbitrary  \textit{directional} derivatives,
\begin{equation}\label{fv}
{\displaystyle {\bf f}_{{\bf v}_i}({\bf n}) \equiv \left. \frac{\partial {\bf f}({\bf n}  + \lambda {\bf v}_i)}
{\partial \lambda }\right \vert_{\lambda = 0} = \left. \frac{\partial {\bf q}[{\bf n}  + \lambda {\bf v}_i]}
{\partial \lambda }\right \vert_{\lambda = 0} - {\bf v}_i}~,
\end{equation}
where the directional derivatives can be calculated using quantum perturbation theory as described in Refs.\ \cite{ANiklasson20,ANiklasson20b}.
It is easy to show that ${\bf f}_{{\bf v}_k}({\bf n}) = {\bf J} {\bf v}_k$. We then introduce the notation
\begin{align}
&{\widetilde {\bf f}}_{{\bf v}_k}({\bf n}) \equiv  ~{\bf K}_0 {\bf f}_{{\bf v}_k}({\bf n})\\
&{\widetilde {\bf f}}{\bf (n)} =~ {\bf K}_0  {\bf f(n)} = {\bf K}_0 \left({\bf q}[{\bf n}] - {\bf n} \right).
\end{align}
The generalize preconditioned Jacobian, ${\bf K}_0{\bf J}$, can then be expressed as
\begin{align}
{\bf K}_0{\bf J} = \sum_{k,l = 1}^N{\widetilde {\bf f}_{{\bf v}_k}} L_{kl} {\bf v}_l^{\rm T}.
\end{align}
Here ${\bf L} = {\bf O}^{-1}$, where $O_{ij} = {\bf v}_i^T{\bf v}_j$. 
The corresponding inverse, i.e.\ the preconditioned kernel, is then 
\begin{align}
\left({\bf K}_0{\bf J}\right)^{-1} = \sum_{k,l=1}^N {\bf v}_k {M_{kl}} {\widetilde {\bf f}_{{\bf v}_l}}^T,
\end{align}
where ${{\bf M}} = {\bf S}^{-1}$, with $S_{ij} = {\widetilde {\bf f}}_i^T{\widetilde {\bf f}}_j$.

If $\left({\bf K}_0{\bf J}\right) \approx {\bf I}$ we can in general use a low-rank (rank-$m$) approximation for how the preconditioned kernel acts on the preconditioned residual vector in Eq.\ (\ref{Eq_Mot3}). The rank-$m$ kernel approximation is then given by
\begin{align}\label{mIJacobian}
\left({\bf K}_0{\bf J}\right)^{-1} = \sum_{k,l=1}^m {\bf v}_k {M_{kl}} {\widetilde {\bf f}_{{\bf v}_l}}^T,~~ m \le N,
\end{align}
where we chose the vectors, $\{{\bf v}_l\}$, from the preconditioned, orthogonalized Krylov subspace expansion,
\begin{align}
\left\{{\bf v}_k\right\} \in {\rm span}^\perp \left\{ {\widetilde {\bf f}}({\bf n}), ({\bf K}_0 {\bf J}) {\widetilde {\bf f}}({\bf n}),({\bf K}_0 {\bf J})^2 {\widetilde {\bf f}}({\bf n}), \ldots  \right\}.
\end{align}
The preconditioned Krylov subspace approximation of the kernel in Eq.\ (\ref{mIJacobian}) can then be used in the integration of the equations of motion for the electronic degrees of freedom in Eq.\ (\ref{Eq_Mot3}) as described in Ref.\ \cite{ANiklasson20}.

\section{sXL-BOMD simulations using SCC-DFTB theory}

To demonstrate the sXL-BOMD scheme presented in this article we will use SCC-DFTB theory
\cite{MElstner98,MFinnis98,TFrauenheim00,MGaus11,BAradi15,BHourahine20} as implemented in a spin-polarized version of the open-source electronic structure software package LATTE \cite{LATTE,MCawkwell12,AKrishnapriyan17}. Spin-polarized SCC-DFTB is an approximation of first-principles Kohn-Sham density functional theory. It is derived from a second or third-order expansion of the Kohn-Sham energy density functional in the charge and spin fluctuations around a set of overlapping atomic spin and charge densities. 

SCC-DFTB theory can be used to demonstrate the most important features of the sXL-BOMD scheme, in particular, how we can avoid the iterative SCF optimization process while still generating stable molecular trajectories. First we present the particular spin-polarized SCC-DFTB shadow energy functional and Born-Oppenheimer potential. We then present a QMD simulation example for a supercell of bcc Fe. We also show how the preconditioned Krylov subspace approximation of the kernel can be used to accelerate the convergence of the regular SCF optimization. Thereafter we show how the parameterization of the SCC-DFTB energy expression was performed based on first-principles DFT.

\subsection{SCC-DFTB formulation}

The SCC-DFTB framework \cite{MElstner98,MFinnis98,TFrauenheim00,MGaus11,BAradi15,BHourahine20} provides a natural description of the formation of bonds and includes
charge transfer between species of different electronegativities. The electronic free energy is given by
\begin{align}\label{ELatte}
E({\bf R},D) &= \sum_\sigma {\rm Tr}\left[H^\sigma_0(D^\sigma-D^\sigma_0)\right] 
+ \frac{1}{2} \sum_{I,J}^N q_I \gamma_{IJ} q_J \nonumber \\
&+  \frac{1}{2} \sum_I \sum_{l} m^{(I)}_l J^{(I)}_{l} m^{(I)}_{l} 
- \sum_\sigma T_e {\cal S}[D^\sigma],
\end{align}
where $H^\sigma_0$ is the charge-independent 
Slater-Koster tight-binding Hamiltonian that
represents the distance and angularly dependent overlap between valence orbitals
on neighboring atoms for the spin up $\sigma = \uparrow$ and spin-down, $\sigma = \downarrow$ channel. 
$D = \{D^\sigma\}$ is the effective single-particle spin density matrix, $D_0 = \{D^\sigma_0\}$ is the spin density matrix 
for neutral, noninteracting atoms, $\gamma_{IJ}$ is a screened Coulomb potential between overlapping atom-centered Gaussian-like charge distributions, which is equal to the Hubbard-U term for the on-site $\gamma_{II} = U_I$ elements when $I = J$ and decays as $1/|{\bf R}_I - {\bf R}_J|$ at larger distances, $N$ is the number of atoms,  $q_I$
is the Mulliken partial electron occupation on atom I,
\begin{equation}
    q_I  = \sum_\sigma \sum_{i \in I} \sum_j \left(D^\sigma_{ij}-{D^\sigma_0}_{ij}\right)S_{ji}.
\end{equation}
The $\{i\}$ (or $\{j\}$) indices are multi-indices for the orbitals centered at atomic positions, $\{{\bf R}_I\}$ (or $\{{\bf R}_J\}$), with the spherical harmonics $l$ and $m$ labels.
The $l$-dependent net Mulliken spin moments, $m^{(I)}_l$, for atom $I$ are given by
\begin{equation}
    m^{(I)}_l  =    \sum_{i\in I,i \in l} \sum_j \left((D^\uparrow_{ij}-{D^\uparrow_0}_{ij})S_{ji} - S_{ij}(D^\downarrow_{ji}-{D^\downarrow_0}_{ji})\right).
\end{equation}
The last term, $T_e {\cal S}[D^\sigma]$, in Eq.\ (\ref{ELatte}) is the electronic temperature, $T_e$, times the entropy term,
\begin{equation}\label{Entropy}
    {\cal S}[D^\sigma] = {\rm Tr}\left[ D_\sigma^\perp \ln D_\sigma^\perp + (I-D_\sigma^\perp) \ln((I-D_\sigma^\perp) \right],
\end{equation}
expressed in an orthogonal density matrix representation, $D_\sigma^\perp$, where $D^\sigma = Z D_\sigma^\perp Z^T$.  
As before, here $Z$ is the inverse factorization of the overlap matrix, where $Z^TSZ = I$, and where the overlap matrix $S$ has matrix elements $S_{ij} = \langle \phi_i\vert \phi_j\rangle$ for a minimal set of 
atomic orbital basis functions, $\{\phi_i({\bf r})\}$. Finally,  $J^{(I)}_l$ is an $l$-dependent Stoner parameter for the spin-dependent energy term that has been optimized from first principles data.
The Born-Oppenheimer potential energy surface is then given by the constrained minimization
\begin{equation}\label{directBO}
    U({\bf R}) = \min_{D^\sigma} \left\{ E({\bf R},D) \left \vert \sum_\sigma {\rm Tr}[D^\sigma S] = N_e \right.  \right\} + E_{\rm rep}({\bf R}).
\end{equation}
Here $E_{\rm rep}({\bf R})$ is a charge independent energy term that provides strong repulsion at short interatomic distances, and $N_e$ is the total number of valence electrons.

The corresponding shadow Born-Oppenheimer potential is given from the constrained minimization of a shadow energy function, 
which is given from a linearization of $E({\bf R},D)$ in Eq.\ (\ref{ELatte}) around an approximate spin-dependent density matrix, $P = \{P^\sigma\}$, with the spin-resolved partial charge occupations,
\begin{equation} \label{nfromP}
    n_{I,l}^\sigma  =  \sum_{i \in I, i\in l} \sum_j \left(P^\sigma_{ij}-{P^\sigma_0}_{ij}\right)S_{ji}.
\end{equation}
and corresponding spin moments,
\begin{equation}
    s^{(I)}_l  =   n_{I,l}^\uparrow - n_{I,l}^\downarrow,
\end{equation}
and combined net partial charge occupations,
\begin{equation}
    {n}_I  = \sum_l n_{I,l}^\uparrow + n_{I,l}^\downarrow.
\end{equation}
The linearized shadow energy function is then
\begin{align} \label{ShadowELatte}
& {\cal E}({\bf R},D,P)  = \sum_\sigma {\rm Tr}\left[H^\sigma_0(D^\sigma-D^\sigma_0)\right] \nonumber \\
& ~~~~ + \frac{1}{2} \sum_{I,J}^N (2q_I-{ n}_I) \gamma_{IJ} {n}_J \nonumber \\
& ~~~~~~~ + \frac{1}{2} \sum_I \sum_{l,l'} (2 m^{(I)}_l -s^{(I)}_l) J^{(I)}_{l} s^{(I)}_{l} \nonumber \\ 
& ~~~~~~~~~~ - \sum_\sigma T_e {\cal S}[D^\sigma],
\end{align}
which gives us the optimized shadow Born-Oppenheimer potential,
\begin{align}\label{ShadowULatte}
    {\cal U}({\bf R},{\bf n}) = &\min_{D^\sigma} \left\{ {\cal E}({\bf R},D,P)  \left \vert \sum_\sigma {\rm Tr}[D^\sigma S] = N_e \right. \right\} \nonumber \\
&+ E_{\rm rep}({\bf R}).
\end{align}
This optimization can be performed in a single direct step without requiring any iterative SCF optimization. Only a single diagonalization of 
the effective single-particle Kohn-Sham Hamiltonian is necessary. This is in contrast to the regular direct method in 
Eq.\ (\ref{directBO}).  
Notice, that the $P$ dependency in the shadow potential, ${\cal U}$, is replaced by the spin resolved partial Mulliken occupations, ${\bf n} \equiv {\bf n}(P)$, (as in Eq.\ (\ref{TwoComp})) that are given by $P$ in Eq.\ (\ref{nfromP}). We can do this because the partial spin occupations, ${\bf n}$, together with the atomic positions, ${\bf R}$, uniquely determine the value of the shadow potential, ${\cal U}({\bf R},{\bf n})$.

The shadow energy function and shadow Born-Oppenheimer potential in Eqs.\ (\ref{ShadowELatte}) and (\ref{ShadowULatte}) 
are the SCC-DFTB formulations of the corresponding Kohn-Sham expressions in Eqs.\ (\ref{FreeShadowFunc}) and (\ref{ShadowU}).

If we use this shadow Born-Oppenheimer potential in the extended Lagrangian, Eq.\ (\ref{SXL-BOMD}), we get the equations of motion, as in Eq.\ (\ref{EqMot}). The SCC-DFTB theory then follows the same sXL-BOMD formalism for QMD simulations as presented in Section \ref{sXLBOMD}.

\begin{figure}[htb]
\centering
\includegraphics[width=0.5\textwidth]{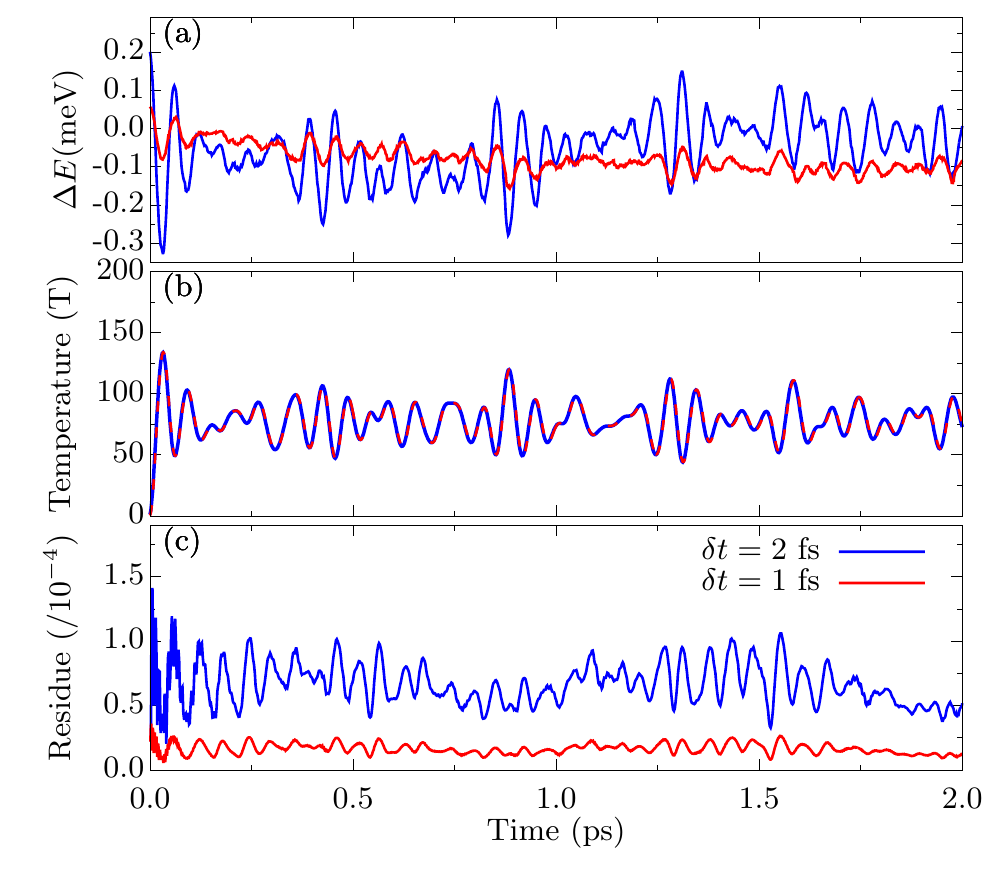}
\caption{\label{Fig:energy}
{\small Spin-polarized XL-BOMD (sXL-BOMD) simulations of Fe supercell using the LATTE electronic structure code based on SCC-DFTB theory. a) shows the fluctuations 
in the total energy per atom, $\Delta E$, b) plots the fluctuations in the statistical temperature and c) is the residue of the spin-resolved charge density given by the root mean square of the residual function ${\bf f}[{\bf n}] = {\bf q}[{\bf n}]- {\bf n}$. 
The sXL-BOMD simulations avoid any iterative SCF-optimization while the fluctuations in the total energy per atom, $\Delta E$, remains stable with no visible systematic drift. 
The low-rank preconditioned Krylov subspace approximation in Eq.\ (\ref{mIJacobian}) was used 
for the kernel, ${\bf K}_{\sigma, \sigma'}$ in the integration of the extended electronic degrees
of freedom, Eq.\ (\ref{Eq_Mot2}).}}
\end{figure}

\begin{figure}[htb]
\centering
\includegraphics[width=0.5\textwidth]{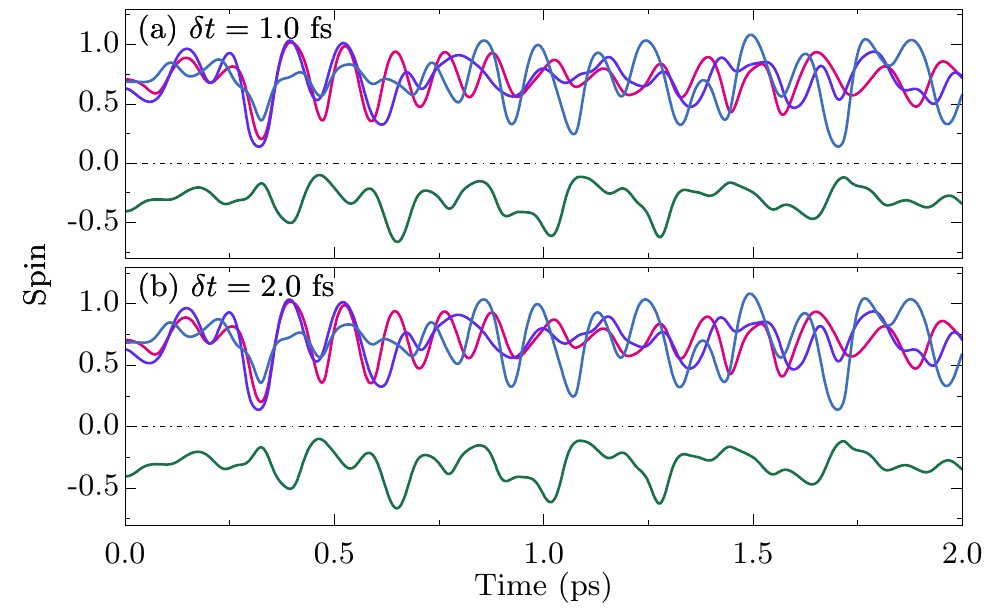}
\caption{\label{Fig:spin}
{\small Spin-polarized XL-BOMD (sXL-BOMD) simulations of Fe supercell using the LATTE electronic structure code based on SCC-DFTB theory. 
a) and b) panels show  the selected set of atom projected spin moments for $\delta t=1$~fs and 2~fs, respectively.  The different colors represent different atoms.
}}
\end{figure}

\subsection{Fe QMD simulation}

For the demonstration of sXL-BOMD we will perform a QMD simulation of ferromagnetic Fe set up in a bcc-like 16 atom supercell with periodic boundary conditions. An integration time steps of 2 and 1 fs was chosen with a statistical temperature oscillating around 100 K. 
The results with the two time steps are shown in Fig.~\ref{Fig:energy}. 
Fig.~\ref{Fig:energy}(a)-(c) shows the fluctuations in the total energy per atom, the fluctuations in the statistical temperature, and the residues (given by the Frobenius norm of the spin orbital-resolved charge density), respectively. The fluctuations in the spin moment of five individual Fe atoms are show in Fig.~\ref{Fig:spin}. As shown by the comparison of the two time steps in Fig.~\ref{Fig:energy}, the size of the residual and energy fluctuation are both reduced by a factor of 4 as the integration time step is halved, demonstrating the approximate $\delta t^2$ scaling of the residual and energy errors.

\begin{figure}
\centering
\includegraphics[scale=.70]{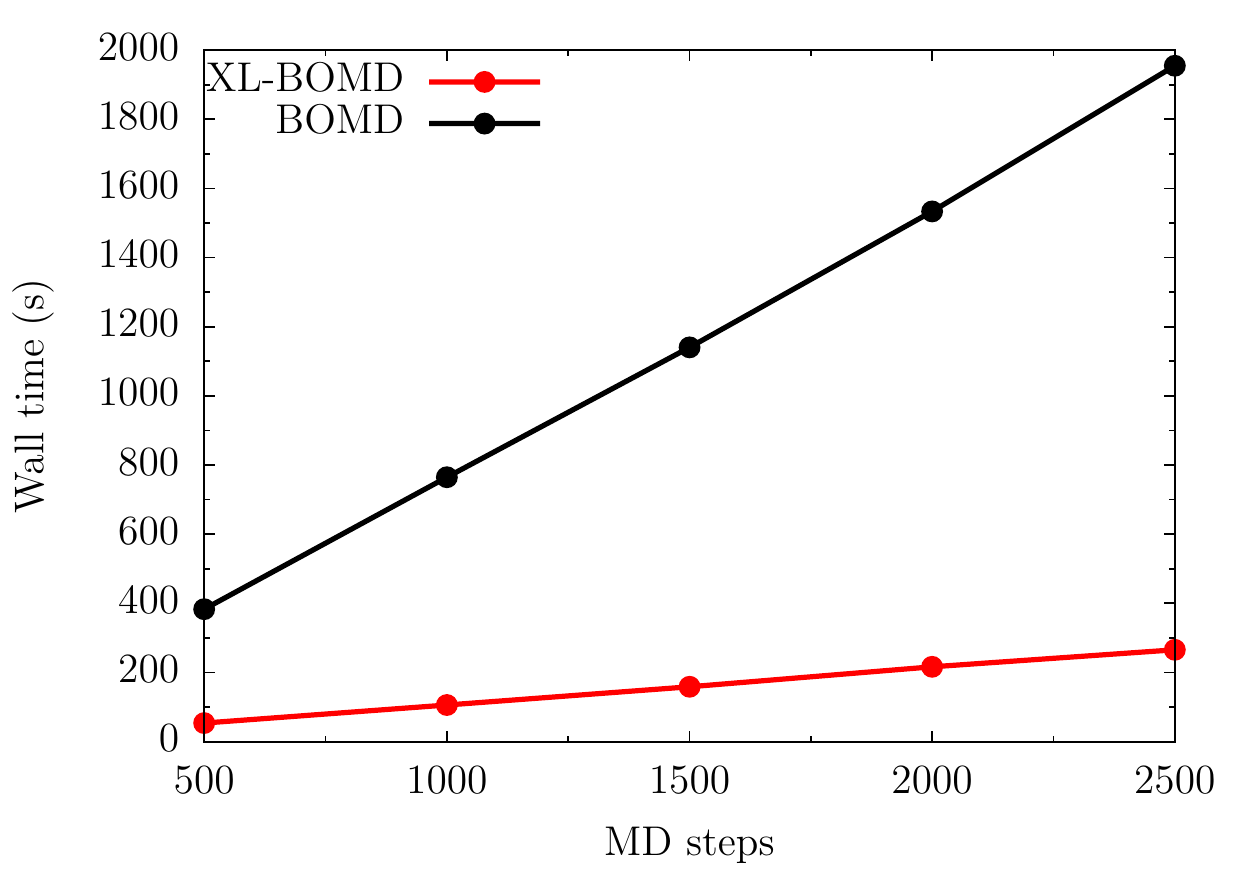}
\caption{\label{Fig_3}
{\small Wall times of spin-polarized XL-BOMD and regular direct spin-polarized BOMD simulations of 16-atom Fe supercell based on SCC-DFTB theory. The XL-BOMD avoids the expensive SCF calculations in each MD step, making it computationally more efficient even compared to a regular BOMD simulation with a loose SCF convergence.}}
\end{figure}

The stability of the MD trajectories can be gauged by the behavior of the total energy, i.e.\ the constant of motion, 
which remains stable without any visible systematic drift, as seen in Fig.~\ref{Fig:energy} a. 
As is evident from Fig.\ \ref{Fig_3}, the computational cost is also significantly reduced 
compared to a regular spin-polarized SCC-DFTB calculation, where the spin-charge density is optimized in each time step. 
Figure \ref{Fig_3} shows the wall clock time of our sXL-BOMD implementation in comparison to a regular 
BOMD simulation with a time-step of 1~fs. For the sXL-BOMD, no iterative SCF optimization 
is needed except in the initial step for which also a kernel preconditioner is calculated. In contrast, regular BOMD requires an 
iterative SCF optimization procedure prior to the force evaluation in each MD step.  
Our SCF optimization uses Pulay's DIIS method \cite{PPulay80,PPulay82}, but still often requires a large number of 
iterations in each MD step to reach even a low convergence threshold ($10^{-4}$ root mean square error (RMSE) in charge and spin), 
resulting in a much larger computational time (black line) compared to the XL-BOMD simulation (red line). As is shown in Fig.~\ref{Fig:energy} the sXL-BOMD scheme is within an even tighter tolerance (for $\delta t = 1$ fs) as estimated from the residue in the charges measured by the same RMSE. In this way the comparison may be somewhat misleading, because the accuracy of the XL-BOMD simulation is in practice higher, especially its global integrated error is much smaller. Our new implementation of a spin-polarized XL-BOMD is simply a major improvement in stability, accuracy, and speed.

\subsection{SCF acceleration}

The preconditioned Krylov subspace approximation of the kernel in Sec.\ \ref{Kernel} can also be used to accelerate the convergence of a regular iterative SCF optimization. This can be achieve with a quasi-Newton scheme, 
\begin{equation}\label{QN}
    {\bf n}_{\rm new} = {\bf n}_{\rm old} - {\bf K} \left( {\bf q} [{\bf n}_{\rm old}] - {\bf n}_{\rm old} \right),
\end{equation}
where the kernel ${\bf K}$ is replaced by the preconditioned Krylov subsapce approximation in Eq.\ (\ref{mIJacobian}).

\begin{figure}[htb]
\centering
\includegraphics[scale=.75]{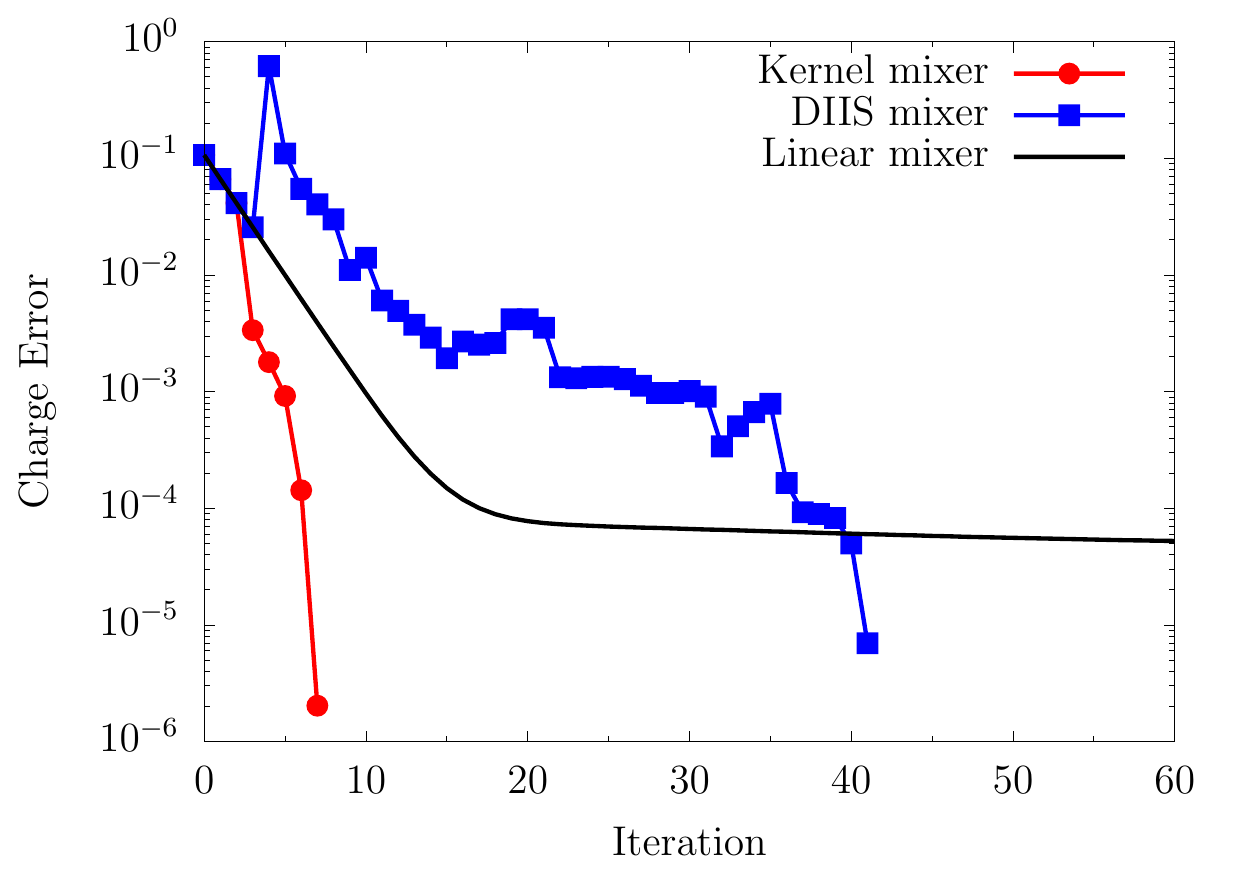}
\caption{\label{Fig_SCF}
{\small Comparison of SCF convergences for magnetic Fe as measured by the root mean square of the residual charge (Charge Error) using Kernel (red), DIIS mixer (blue), and linear mixers (16 atom Fe supercell).
The mixing coefficient  of the linear mixer is 0.06, which is manually tuned to have the best convergence; DIIS and kernel mixers start with linear mixing at the very beginning and are activated when the charge error is smaller than 0.05. Hence, the charge errors of the first 3 SCF iterations are the same. But the kernel mixer significantly accelerates the SCF convergence once it is activated. 
}
}
\end{figure}

Figure \ref{Fig_SCF} shows the SCF convergence as a function of the iteration step using either a simple hand-tuned linear mixer, the DIIS algorithm \cite{PPulay80,PPulay82,SAmatrya16}, or the kernel mixer given by the quasi-Newton scheme in Eq.\ (\ref{QN}). The kernel mixer provides a rapid and stable convergence. However, the cost per iteration is higher. First of all, the kernel mixing included a preconditioner calculated after the charge error with the linear mixer becomes smaller than 0.05. However, because the preconditioner can be reused in the sXL-BOMD integrations scheme, possibly over thousands of time steps, the overhead of calculating the preconditioner can be ignored. The extra cost is then mainly from the low-rank approximations. Here we used 2-3 ranks adaptively in each SCF step \cite{ANiklasson20}.

\subsection{SCC-DFTB parameterization of Fe}

The SCC-DFTB energy expression was parameterized for metallic Fe in a two-step process where 
i) bond and overlap integrals are extracted from non-self consistent Kohn-Sham DFT calculations, 
and ii) the pairwise terms that provide strong repulsion at short interatomic distances are constructed 
so that the complete model reproduces DFT binding energy curves. 
The model for metallic Fe includes itinerant electron ferromagnetism and also allows for an 
alternative route to charge self-consistency, where long-range electrostatic interactions apart 
from the onsite Hubbard-U term can be turned off in Eq.\ (\ref{ELatte}). This should have little 
effect on simulations with only a single element and little interatomic net charge transfer, e.g. the Fe simulation. 
However, in the simulations presented here in Fig.\ \ref{Fig:energy} we have not used this new feature. 
Instead all simulations are performed with the long-range electrostatic interactions included, 
which is valid for a broader range of materials, including systems with significant interatomic charge transfer.

\begin{figure}[htb]
\centering
\includegraphics[scale=.75]{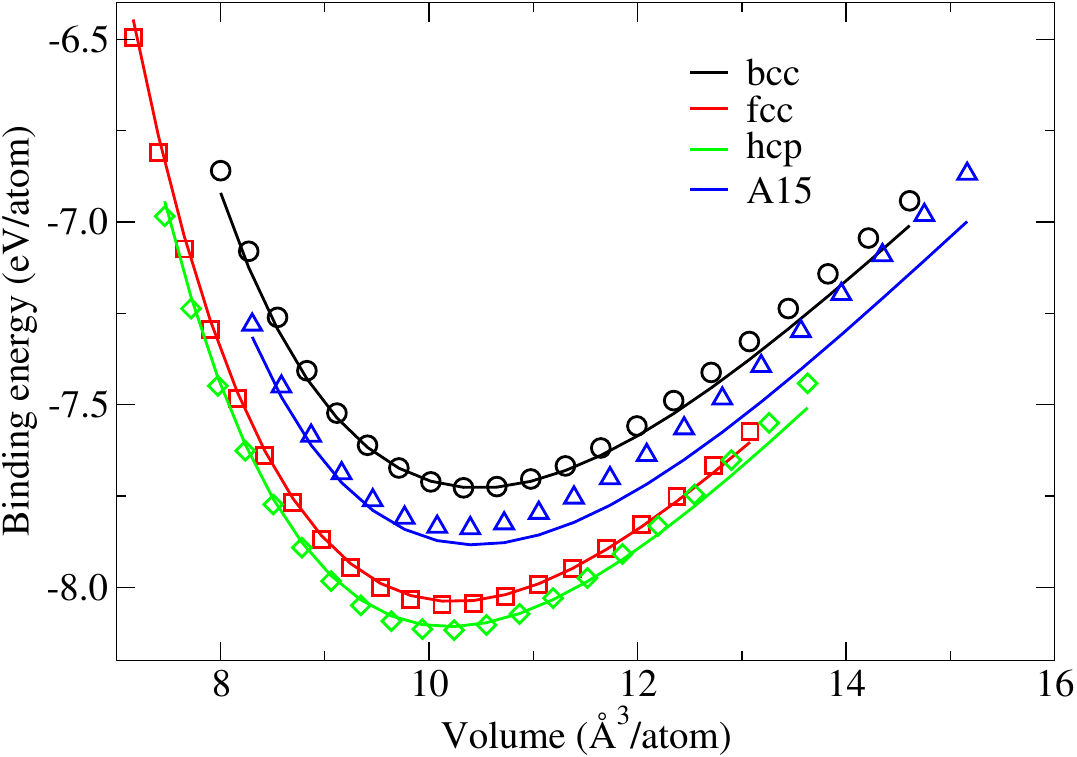}
\caption{\label{Fig_binding}
{\small Binding energy curves for non-magnetic hcp, fcc, bcc, and A15 iron from DFT (symbols) and the optimized SCC-DFTB results (solid lines).}}
\end{figure}

Our SCC-DFTB energy expression for Fe, uses a non-orthogonal basis of s, p, and d orbitals. The bond and overlap integrals, ss${\sigma}$
through dd${\delta}$ were computed from minimal basis DFT calculations using atom-centered numerical orbitals via the PLATO code \cite{SKenny09}. 
The bond and overlap integrals were tabulated for use in LATTE and were not optimized further during the construction of the TB model. 
The repulsive energy term, $E_{\rm rep}$, is approximated as a pairwise term,
\begin{equation}
    E_{\rm rep}=\frac{1}{2}\sum^N_{i\neq j} \Phi(R_{ij}).
\end{equation}
where $\Phi$ is a pair potential and $R_{ij}$ the distance between atoms $i$ and $j$, was constructed so that the SCC-DFTB model reproduced, 
in a least squares sense, binding energy curves for the bcc, fcc, hcp, A15 and simple cubic crystal structures computed using first principles DFT. 
The resulting binding energy curves, in the absence of magnetism, are presented in Fig.~\ref{Fig_binding}.

\section{Summary and Discussion}

We have presented a spin-polarized generalization of XL-BOMD where also the electronic spin-degrees of freedom are included as extended dynamical variables along the atomic positions and their velocities. The new framework looks similar to the original XL-BOMD for non-magnetic or restricted systems, but the number of degrees of freedom doubles. 
This leads to an increased computational cost for the kernel calculation, which includes non-diagonal spin-blocks that couple the response between spin channels. In general, the increased number of degrees of freedom may also lead to additional convergence problems. 

QMD simulations of a Fe supercell with the proposed sXL-BOMD scheme demonstrated stability and a low residual error without relying on any costly iterative SCF optimization prior to the force evaluations. We also showed how the generalized spin-polarized version of the preconditioned Krylov subspace approximation of the kernel could be used in a quasi-Newton scheme to accelerate the convergence of a regular SCF optimization. Such a ground state optimization is also needed in sXL-BOMD simulations, but only in the first initial time step to determine the initial values of ${\bf n}$. The kernel-based quasi-Newton mixer provides a competitive alternative to established SCF acceleration schemes such as the DIIS method \cite{ANiklasson20,VGavini22}.

In QMD simulations, changes in collinear spin configurations, such as spin flips, may occur instantaneously, leading to discontinuities in the potential. Various strategies, such as higher electronic temperatures, adaptive time stepping, or additional constraints, can possibly mitigate this issue. However, a more effective solution requires a non-collinear spin dynamics, which evolves on a timescale similar to the nuclear degrees of freedom. This involves significant modifications to the current sXL-BOMD model, and is a direction for future research.

%

\section{Acknowledgements}

This work is supported by the U.S. Department of Energy Office of Basic Energy Sciences (FWP LANLE8AN) and by the U.S. Department of Energy through the Los Alamos National Laboratory. This research was also supported by the Exascale Computing Project (17-SC-20-SC), a collaborative effort of the U.S. Department of Energy Office of Science and the National Nuclear Security Administration. M.J.C. and R.P. were supported by the eXtremeMAT program of the US Department of Energy Office of Fossil Energy and thank Laurent Capolungo for many useful discussions.  Los Alamos National Laboratory is operated by Triad National Security, LLC, for the National Nuclear Security Administration of the U.S. Department of Energy Contract No. 892333218NCA000001.

\bibliography{mndo_new_xy}

\end{document}